\documentclass[usenatbib]{mn2e}

\usepackage{url}
\usepackage{graphicx}
\usepackage{multirow}

\newcommand{\graph}[2][82 80 530 670]
{\includegraphics[angle=90, width=\columnwidth, bb=#1]{#2}}

\newcommand{\graphtwo}[2][20 80 567 680]
{\includegraphics[angle=90, width=\columnwidth, bb=#1]{#2}}

\newcommand{\graphfour}[2][25 73 550 695]
{\includegraphics[angle=90, width=\columnwidth, bb=#1]{#2}}

\begin{document}

\title{Pre-main-sequence variability across 
       the radiative-convective gap}
 
   \author[Saunders et al.]
   {Eric S. Saunders$^{1,2}$ \thanks{e-mail: esaunders@lcogt.net},
    Tim Naylor$^{2}$, Nathan Mayne$^{2}$ and S.P. Littlefair$^{3}$    
   \\ $^{1}$ Las Cumbres Observatory, 6740 Cortona Dr., Suite 102, 
   Santa Barbara, CA 93117
   \\ $^{2}$ School of Physics, University of Exeter, Stocker Road, Exeter EX4
   4QL
   \\ $^{3}$ Department of Physics and Astronomy, University of Sheffield,
   Sheffield S3 7RH
   }
   
\maketitle

\begin{abstract}

  We use $I$ band imaging to perform a variability survey of the 13\,Myr-old
cluster h Per. We find a significant fraction of the cluster members to be
variable. Most importantly, we find that variable members lie almost entirely on
the convective side of the gap in the cluster sequence between fully convective
stars and those which have a radiative core. This result is consistent with a
scenario in which the magnetic field changes topology when the star changes from
being fully convective, to one containing a radiative core.  When the star is
convective the magnetic field appears dominated by large-scale structures,
resulting in global-size spots that drive the observed variability. For those
stars with radiative cores we observe a marked absence of variability due to
spots, which suggests a switch to a magnetic field dominated by  smaller-scale
structures, resulting in many smaller spots and thus less apparent variability. 
This implies that wide field variability surveys may only be sensitive to fully
convective stars.  On the one hand this reduces the chances of picking out young
groups (since the convective stars are the lower mass and therefore fainter
objects), but conversely the absolute magnitude of the head of the convective
sequence provides a straightforward measure of age for those groups which are
discovered. \end{abstract}


\section{Introduction} \label{Section:Intro}
   
Despite the widespread use of the photometric variability of young
stars to study their rotational evolution \citep[see][for a
review]{2007prpl.conf..297H}, its underlying causes remain poorly
understood.  The variability itself appears to be due to a combination
of (in order of importance) cool spots, hot spots, obscuration and
pulsation
\citep{1999A&A...345..505K,2001AJ....121.3160C,2007A&A...461..183G,2008A&A...479..827G}.
We know these stars have kG magnetic fields
\citep[e.g.][]{2007ApJ...664..975J}, and all these sources of
variability are probably related to those fields.  The cool spots are
large starspots, the hotspots are probably due to accretion funnelled
onto a small region of the star by the magnetic field, and the
obscuration is again due to magnetically confined accretion
structures. A largely unexploited route to elucidate the underlying
physics of the resulting optical variability is to study its
characteristics as a function of fundamental stellar parameters such
as magnetic field, age and accretion rate. Variability caused by
pulsation will be of order millimagnitudes and so difficult to
observe. Additionally, variability caused by obscuration due to
circumstellar, or disc, material is expected to be negligible for h
Per, as at $\approx$ 13 Myrs
\citep{Slesnick02persei,mayne07_empirical_isochrones} most of the
associated discs and circumstellar material would probably have
dissipated. However, variability caused by large cool spots or
accretion hotspots generates significant magnitude changes over
periods of hours to days, and therefore is relatively simple to
detect.  Here we study the degree of variability, caused by hot and
cool spots, as a function of stellar mass in the young cluster h
Per. The mass range covered by our data includes stars either side of
the transition from low-mass fully convective to stars with radiative
cores (the point in the colour-magnitude diagram
\cite{mayne07_empirical_isochrones} call the radiative-convective
gap).

   %% Age of cluster. Background info. Previous work.
   
The target of our investigation, h Per, and its companion $\chi$ Per
are young, bright open clusters that have been the subject of detailed
observations for seventy years \citep[e.g.,][]{Oosterhoff37persei,
  Crawford70persei, Tapia84persei, Waelkens90persei}. More recently,
attention has been focused on their mass functions and mass
segregation \citep{Slesnick02persei, Bragg05persei}. We adopt an age
of 13 Myr \citep{Slesnick02persei,mayne07_empirical_isochrones},
although the precise value is not crucial for our work.

   The remainder of this paper is laid out as follows.  In Section
   \ref{Section:Method} we describe the data acquisition and reduction
   processes, and the selection criteria used to identify
   variability. In Section \ref{Section:Results} we present a
   colour-magnitude diagram (CMD) of the variable stars, alongside a
   brief statistical analysis of the gap detection significance. In
   Section \ref{Section:Discussion} we discuss our findings and their
   implications. Finally, Section \ref{Section:Conclusion} is a short
   conclusion.

\section{Method} \label{Section:Method}

   %% Details of dataset. 
   
   The observations were taken using a Sloan \emph{i} filter with the 2.5m
   Isaac Newton Telescope (INT) on La Palma, using the Wide Field Camera with
   four EEV CCDs. The run spanned 16 nights, from the 22nd September
   to the 7th of October, 2004. 
   h Per was observed on 12 of these nights, for
   around  2.5 hours per night. Exposures of 5, 30 and 300 seconds were
   obtained. There were 213 good frames taken with 300s exposure, and
   110 good frames for each of 5\,s and 30\,s exposure.

\subsection{Data reduction} \label{Section:Data reduction}
   
   %% Summary of data reduction...
   %% Details of initial data reduction stages - bias, defringing, flat-fielding
   %% Details of object detection algorithm...
   %% Details of the astrometric solution...

The data were reduced, and then photometry extracted using the optimal
extraction algorithm of \citet{Naylor98OptimalExtraction} and
\citet{Naylor02OptimalExtraction}, with the modifications introduced
by \citet{Littlefair05IC348}. The images were bias subtracted using a
single master bias frame, corrected for flat-field effects using
twilight sky flat-field images, and defringed. For each exposure time
we chose a single frame, and then searched the combination of the
three to create a master catalogue. The objects were then matched to a
2MASS catalogue of our field to produce an astrometric solution, with
a mean rms discrepency in position of 0.1 arcseconds.

   %% Application of optimal photometry to the dataset...
   %% Calculation of offsets...
   
   For each image positional and rotational offsets were determined with respect
   to the master catalogue. Using the list of candidate objects we performed
   optimal photometry on individual frames to produce fluxes for each star in
   each image.

   %% Explanation of profile correction...

   Next a ``profile correction'', the logical analogue of an aperture correction
   in standard aperture photometry, was  applied to correct the flux measured
   for our objects to a somewhat larger aperture (although the crowded nature of
   the field limits this). 
   A number of bright, unsaturated stars in each CCD provided
   the profile correction for that CCD, which is a polynomial function of
   position. 
   Although our photometry is relative, and not tied to any particular
   system, this correction is still important because the spatial 
   variation of the PSF is a function of the seeing \citep{Littlefair05IC348}.

   %% Normalisation of frames for transparency correction...
   
   From this point we separated the photometry by exposure time into three
   datasets. For each dataset a relative transparency correction was applied to
   all  the resulting photometric measurements, to normalise any differences
   between frames arising from variations in airmass or transparency. This was
   achieved by selecting a subset of relatively bright, unvarying stars and
   using their average magnitude to define a reference brightness for each
   frame. This was done in the same way as described in
   \citet{Littlefair05IC348}.

   %% Discussion of instrumental magnitudes.
   
\begin{figure}  
      \graph{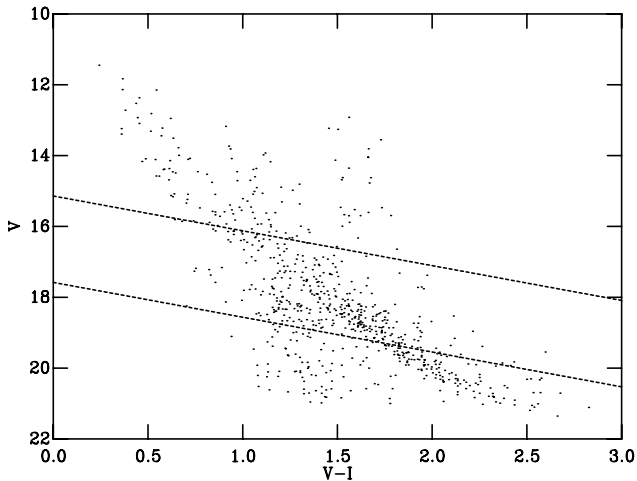}
      
      \caption{CMD of the variable population, after combining the 5\,s, 30\,s
        and 300\,s datasets (724 objects). The dashed lines indicate the
        limiting magnitudes (discussed in Section \ref{Section:Data reduction})
        used to combine the datasets. The fact that the sequence appears
        unbroken across the boundaries means that the selection criteria will
        not be a function of magnitude and therefore
        sensitivity.\label{fig:final_dataset}}

   \end{figure}

   The photometry for a star in each dataset is only relative
   photometry with respect to the other stars in that dataset. To
   produce the final catalogue, the datasets were renormalised using
   the magnitude ranges for which they overlapped. We then applied an
   arbitrary zero point to our instrumental magnitude so that our
   observations would be in a natural magnitude system approximately
   comparable with Sloan \emph{i}. We denote magnitudes in this system
   by the symbol $i_I$. Of course many stars have magnitudes from more
   than one dataset, but for consistency it is important we create
   each lightcurve from a single dataset, and that we use the same
   dataset for stars of similar mean magnitudes. Thus the 5\,s dataset
   was used for all stars brighter than an $i_I$ of 15.2. Lightcurves
   for stars with $15.2 < i_I < 17.5$ were taken from the 30\,s
   dataset, while for stars fainter than $i_I =17.5$ lightcurves were
   taken from the 300\,s dataset. These characteristic or limiting
   magnitudes are shown plotted over the data as dashed lines in
   Figure \ref{fig:final_dataset}.

\subsection{Variable star selection} \label{Section:Variable star selection}
   
   %% Introduce selection mechanism.

   %% Describe removal of flagged stars.
   %% Describe removal of spurious variables in corner of CCD 3.
   %% Choice of S-N cut.

   The identification of variable stars involved a set of distinct selection
   criteria. First, the dataset was stripped of stars flagged as problematic
   during the data reduction process. Data quality checks made during this
   process included identifying duplicate detections of the same star, stars
   containing bad or saturated pixels, pixels with counts in the non-linear
   response range of the detector, and pixels with negative counts.
   Additionally, stars possessing a PSF that was not point-like were flagged for
   non-stellarity. A number of spurious variables at the corner of CCD 3 were
   removed due to high vignetting, using a simple coordinate cut. We then
   selected only those lightcurves with a mean  signal-to-noise per frame of
   more than 10.

   %% Minimum number of good points in lightcurve. Justification.
   
   Having dealt with the most obviously problematic objects, we next
   considered the lightcurves themselves.  The ideal datset would have
   the same number of datapoints for each lightcurve, allowing us to
   use a simple $\chi^2$ cut to identify variable stars. Of course
   many lightcurves have missing datapoints for a variety of reasons,
   which means that we should choose a different value of $\chi^2$ for
   each lightcurve if we require identical probabilities that the
   lightcurves show variability. In practice, for a large number of
   degrees of freedom the distribution of reduced $\chi^2$ is such
   that the probability associated with a given $\chi^2_\nu$ is almost
   independent of $\nu$. This means if we pick only those lightcurves
   with a large number of datapoints, we can apply a simple
   $\chi^2_\nu$ cutoff. Such an approach has the advantage that even
   if our model of the uncertainties is not perfect (in particular
   there may be correlated noise in the data), the cut-off will be
   consistent. So we required a minimum of 60 good datapoints from the
   5 and 30 second datasets (from a possible 110), and 120 good
   datapoints from the 300 second dataset (from a possible 213).  To
   make our analysis more robust against single spurious datapoints we
   discarded the datapoint with the highest individual $\chi^2$ with
   respect to the weighted mean, before re-calculating the mean, and
   hence $\chi^2_\nu$.

   %% Choice of chisq cut.

   Following \citet{Littlefair05IC348}, we chose a threshold of $\chi^2_\nu >
   10$ as a cutoff indicating significant variability in the 300\,s dataset. The
   correct $\chi^2_\nu$ limit for the 30\,s dataset was then determined by
   considering stars falling within the overlap region between the two datasets.
   Varying the value of $\chi^2_\nu$ applied to the 30\,s dataset, we found that
   a threshold of $\chi^2_\nu > 3$ recovered a number of variable stars
   in this region that was comparable to the number of stars detected in the
   300\,s dataset. Similarly, by considering variables identified in both the
   5\,s and 30\,s dataset, it was found that a $\chi^2_\nu > 3$ threshold
   applied to the 5\,s dataset recovered all variables to $i_I \approx 15.2$.
   This process yielded 724 significantly varying objects from the initial list
   of 19860 objects with good quality lightcurves.
   
   %% Details of Nathan's dataset.

\subsection{Periodic Variability}

   %% Present periods for the objects with reasonable lightcurves.

One might hope to reduce the contamination still further by considering only
those stars which have periodic modulation. Although our dataset is not well
suited to identifying periods, caused by magnetically formed spots on the
stellar surface, in the likely range for PMS stars (a few hours to $\approx$20
days), we did carry out an exploratory period search, summarised in Table
\ref{Table:periodic_vars}. The period search was performed following the method
described in \citet{Littlefair05IC348}. Final analysis of the shortlist of
variables was made by visual inspection of individual lightcurves. Here we show
our best candidates for periodic variability, along with a note of their
position in the CMD. The number of periodic variables is clearly too small for
our analysis, and so we instead use the catalogue of variable stars to create
colour-magnitude diagrams.

   \begin{table*}
      \begin{tabular}{rrrrrrrr}
         \hline
         ID & RA & Dec & $V$ & $V$ uncertainty & $V$-$I$ & $V$-$I$ uncertainty &
         Period (days)\\
         \hline
1.01 1293 & 02 20 44.582 & +56 54 56.35 & 18.948 & 0.013 &  1.935  & 0.017 & 0.1186993\\
1.02 1325 & 02 17 46.554 & +57 04 18.08 & 18.312 & 0.012 &  1.337  & 0.019 & 0.1464187\\
1.02 2157 & 02 18 01.071 & +56 53 56.73 & 18.770 & 0.015 &  1.159  & 0.024 & 0.2099175\\
1.04 0251 & 02 20 21.400 & +57 03 22.03 & 18.685 & 0.010 &  1.994  & 0.014 & 0.1160178\\
1.04 1906 & 02 20 28.387 & +57 03 03.66 & 19.325 & 0.015 &  1.854  & 0.021 & 0.3029911\\
         \hline
      \end{tabular}
      
      \caption{Periods for the five strongest periodic variable candidates in
      our sample. \label{Table:periodic_vars}}

   \end{table*}

\subsection{Colour-magnitude diagrams} \label{Section:CMD}

To create CMDs for the variable stars we cross matched our catalogue
of variable stars against the h and $\chi$ Per catalogue of
\citet{mayne07_empirical_isochrones}. This gave the $V$ vs $V$-$I$ CMD
with 724 objects shown in Fig.\,\ref{fig:final_dataset}. To enhance
the contrast between the cluster and the contaminating stars (see
Section \ref{Section:Results}) we also created CMDs from both our
variable star catalogue and the \citet{mayne07_empirical_isochrones}
catalogues for a 10 arcminute radius region centered on the cluster
($\alpha=02\ 18\ 58.76$, $\delta=+57\ 08\ 16.5$ J2000). With this
spatial cut, these catalogues have 230 and 3840 objects, respectively,
plotted in Fig.\,\ref{fig:10arc_cmds}.

\section{Results} \label{Section:Results}   

   %% Describe CMD. Dealing with the problem of contamination.
   
\begin{figure}
      \graphtwo{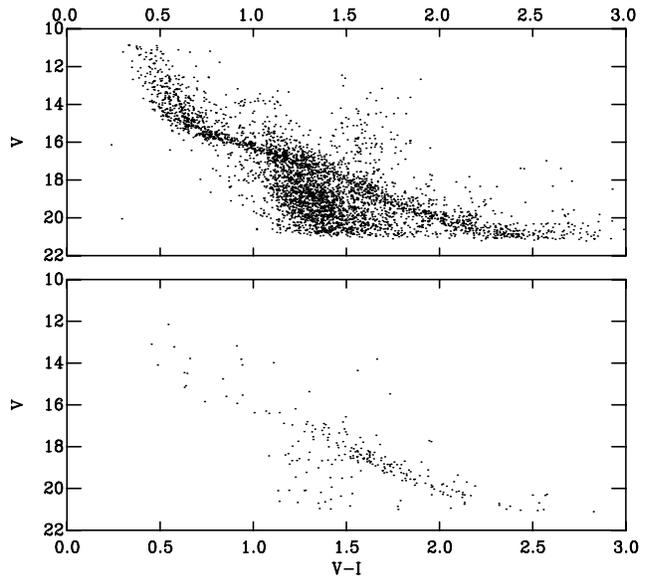}   
      \caption{CMD of photometry towards h Per showing a clear sequence of
        probable members, including variables, taken from
        \citet{mayne07_empirical_isochrones} (\emph{upper panel}), and variables only
        (\emph{lower panel})  after a 10'' radial cut centred on h Per. The upper panel
        has 3840 objects; the lower, 230. The number of variables drops at 
        $V$-$I<1.5$. \label{fig:10arc_cmds}}
   \end{figure}

   Fig.\,\ref{fig:final_dataset} shows a clear pre-main-sequence population
   running diagonally from approximately 14th to 20th magnitude in $V$. However,
   background contamination is also evident. The contamination is more obvious
   in the upper panel of  Fig.\,\ref{fig:10arc_cmds} where it covers a large,
   wedge-shaped area that cuts laterally across the sequence.  The majority of
   this wedge is likely to be dwarfs, but the additional ``finger'' at $V$-$I$
   $\approx 1.6$ is probably due to background giants. As described in Section
   \ref{Section:CMD}, to reduce this contamination we selected only those
   variable stars within 10 arcmin of the cluster core. Fig.\,\ref{fig:10arc_cmds}
   (lower panel) shows the variables remaining after this process. We can
   clearly see that the density of stars falls dramatically as we progress up
   the sequence. This is quite different to the distribution of cluster members
   (Fig.\,\ref{fig:10arc_cmds}, top panel), which shows a narrow sequence up to
   around 15th magnitude.

   %% Plot: Nathan's members after 10 arcminute selection.
   %% Plot: Variable members after 10 arcminute selection.
   
\subsection{Histogram analysis \label{Section:Histogram analysis}}

   %% Motivation. Explain binning: parameters, etc.

\begin{figure}
      \graph{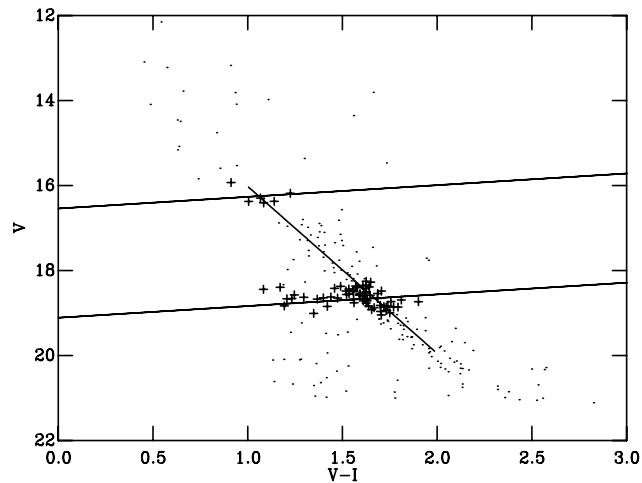}
      \caption{Placement of cuts along the line of the sequence of h Per
               variables. Stars falling within either of the cuts are
               represented as crosses. Identical cuts were used to analyse the
               general cluster membership. \label{fig:bin_members}}
   \end{figure}

\begin{figure}
      \graphfour{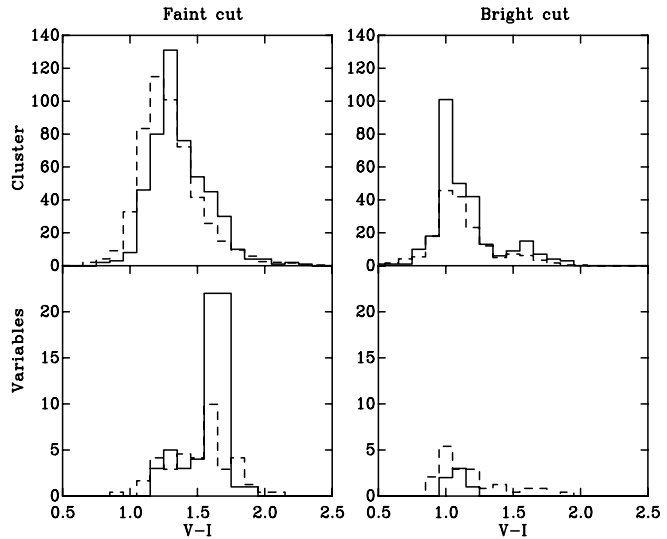}
      
      \caption{Histograms of the lower cut (left hand figures) and upper cut
      (right hand figures), binned along the line of the cut. The upper panels
      show non-variables, while the lower panels present only stars exhibiting
      significant variability. The solid line indicates objects falling within
      a 10' circle of the h Per centre. The dashed line is a count of the stars
      falling outside that cut, normalised to the same area as the cut. The
      dashed lines thus measure the background contamination present in the
      field. \label{fig:low_high_hists}}

   \end{figure}

Although it is clear from visual inspection of the sequence that the
density of variable stars in the CMD falls dramatically at brighter
magnitudes, additional analysis is required to investigate this
behaviour in a robust way.  Because we wish to examine the
distribution of variables along the cluster sequence, we construct
histograms according to the following scheme.  A straight line was
fitted to the central portion of the sequence, running from 1.0 to 2.0
in $V$-$I$. Two lines perpendicular to this sequence line in $V$-$V-I$
space (i.e. appearing perpendicular to the sequence when both axes are
plotted with units of equal size) were then taken, a bright cut
crossing the sequence at 1.0 in $V$-$I$, and a faint cut crossing at
1.66 (see Figure \ref{fig:bin_members}). We selected the stars within
0.4 mags of these lines and then pulled them back onto the cut
centre-lines and binned them in $V$-$I$. Identical cuts were used for
the h Per catalogue of \citet{mayne07_empirical_isochrones}. These
cuts produce the histograms of Figure \ref{fig:low_high_hists}.

   %% Plot: Stars selected for binning.

   %% Plot: Histograms of members, variables.

   %% Discussion of the histograms - low line.

   The histograms show in detail what can be seen qualitatively by eye in the
   CMDs of Fig.\,\ref{fig:10arc_cmds}. While the general cluster membership
   continues to relatively blue colours, the variability dies out blueward of
   about $V$-$I=1.5$. In the bottom-left panel of Fig.\,\ref{fig:low_high_hists},
   which shows the faint magnitude (lower) cut, the variables clearly trace the
   sequence at 1.6--1.7 in colour. The sequence can also be seen in the
   histogram for the general cluster (Fig.\,\ref{fig:low_high_hists}, top-left),
   as a shoulder of excess sources over the background, although in this case
   the contamination dominates between 1.0--1.5 in $V$-$I$. That these stars are
   mostly contaminants is evidenced by the large number of background objects
   present in the sky around the cluster at these colours (dashed line,
   normalised to the same area as the selected area of the cluster). Note that
   the blue end of the contamination appears to be shifted bluewards, relative
   to the foreground cluster. This could be due to differential reddening
   \citep{Capilla02persei}.

   %% Discussion of the histograms - high line.
   
   In the brighter magnitude cut (Fig.\,\ref{fig:low_high_hists}, top-right) the
   cluster is significantly less contaminated, with more than half of the stars
   additional to the background. However, looking at the variables
   (Fig.\,\ref{fig:low_high_hists}, bottom-right) presents a different picture.
   There are relatively few variables present either in the background or in the
   sequence. In fact, the number of variables is consistent with a scenario in
   which there are no variables present in the sequence.

  %% Explain statistical significance

   The position of the two cuts was chosen to bracket the radiative-convective
   gap \citep[as defined and identified in h Per
   by][]{mayne07_empirical_isochrones}, where the stars move from the fully
   convective regime to development of a radiative core. Therefore, these data
   and the chosen cuts allow us to analyse changes in the fraction of variables
   either side of this gap. Integrating under the histograms of Figure
   \ref{fig:low_high_hists} allows us to construct the background-subtracted
   numbers of variable and non-variable stars, and calculate the fraction of
   stars which are variable, both above and below the gap.  We find that below
   the gap 37.9 percent of stars are variables, above it --1.7 percent.  The
   latter is negative as the total (background plus cluster) number of variables
   is so small that its associated noise means that background subtraction can
   push it below zero. This means that we actually see less variables in this
   region of the CMD than the mean number we would expect to be contributed by
   background objects. To test the significance of this difference we formed the
   null hypothesis that the fraction of variables above the gap is the same as
   that below the gap, i.e. 37.9 percent.  We then used the Poisson distribution
   to simulate observations of the stars above the gap, using parent populations
   identical to our observations, save that for cluster stars above the gap we
   changed the fraction of variables to 37.9 percent.  In one million
   simulations we never obtained an observed fraction of variable stars less
   than or equal to --1.7 percent. This allows us to reject the null hypothesis
   that the ratio of members to variable stars is constant either side of the RC
   gap with great confidence, and conclude that the variable stars do indeed die
   out as one moves above the gap.

  \section{Discussion} \label{Section:Discussion}

   \subsection{A break at the RC gap?}   \label{subsection:rc_gap}

   %% Implications for star formation.

  \citet{Stolte04persei} and \citet{mayne07_empirical_isochrones} show that in
  CMDs there is a discontinuity in the stellar sequence between
  pre-main-sequence and main-sequence stars, which
  \citet{mayne07_empirical_isochrones} term the radiative-convective (RC) gap. 
  Below this gap theory predicts that the stars are fully convective, whilst
  above it they possess radiative cores. This gap closes with age, becoming very
  small for objects as old as h Per. For stars above the gap, one model
  postulates that the magnetic field is generated at the tachocline between the
  radiative core and the convective envelope. In this model, it is believed that
  shearing processes in this layer generate a magnetic field, although the full
  picture may well be more complicated \citep[e.g.][]{choudhuri95_solar_dynamo}.
  For fully convective stars the source of the field is more controversial.  On
  the one hand a turbulent velocity field might create a small-scale magnetic
  field \citep{durney93_small_magnetic}. Alternatively
  \citet{chabrier06_large_magnetic} showed that a stratified rotating medium
  might create a large-scale magnetic field (the $\alpha^2$ dynamo).

   \citet{mayne07_empirical_isochrones} identify the position of the RC gap in
   h Per as $V=17.66$ and $V$-$I=1.33$.   It is clear from Fig.\,\ref{fig:10arc_cmds}
   that there is a marked  increase in the  number of variable stars below this
   position in the sequence. This supports the conclusion that the mechanisms
   giving rise to the variability are a function of convective processes in the
   star, and are curtailed by the development of a radiative core. 
   
   One interpretation is that stars below the gap (which are fully convective)
   have large-scale spot structures, as evidenced by their variability, implying
   equivalently large-scale magnetic fields. In contrast, stars above the gap,
   which have a radiative core and greatly reduced convective zone appear to be
   less dominated by large-scale magnetic field structures, as evidenced by the
   dramatic absence of variability arising from spot structures. This implies
   either an absence or sufficient uniformity of distribution of spots,
   resulting in a modulation, if any, below our threshold of detection. Thus our
   observations are consistent with the idea that whilst fully convective stars
   have large-scale structure in their magnetic fields, once the stars develop a
   radiative core the spots become more evenly distributed over the surface.
   Whilst it can still be argued whether the magnetic field is due to 
   convective turnover \citep{chabrier06_large_magnetic},  turbulence
   \citep{durney93_small_magnetic} or fossil field \citep{1987MNRAS.227..553T}, 
   what this study makes clear is that the fully convective structure  is a
   crucial ingredient for the maintenance of the large-scale magnetic fields of
   T Tauri stars.

   \subsection{Comparison with older field
     stars} \label{subsection:field_stars}
  
   Further support for a change in magnetic field topology at the RC
   gap can be found from recent studies using Zeeman Doppler imaging
   of older field dwarfs, where the same physical mechanisms are
   apparent, i.e. fully convective lower mass stars and higher mass
   ojects with radiative cores. These studies have found significant
   evidence for a sharp reduction in magnetic flux, contained within
   large scale structures, as one moves across the boundary (in mass)
   between fully convective stars and those with radiative
   cores. \cite{2008MNRAS.390..545D} and \cite{2008MNRAS.390..567M}
   compare observations of the Stokes V line profiles for old field
   dwarfs about the boundary of radiative core presence \citep[at
   $M_*\approx 0.35M_{\odot}$ using the evolutionary models of
   ][]{1997A&A...327.1039C}, to simulated profiles from estimated
   magnetic maps of the stellar surface. The Stokes V measurements are
   sensitive to large scale magnetic fields, rather than small scale
   structures, which have opposite polarities that cancel in the
   Stokes V profile.  \cite{2008MNRAS.390..545D} and
   \cite{2008MNRAS.390..567M} find a clear difference in magnetic
   fields, with large scale poloidal, axisymmetric fields dominant for
   fully convective stars, and only toriodal and non-axisymmetric
   poloidal fields apparent over large scales for stars with radiative
   cores.  Measurements of X-ray activity and the total magnetic field
   from Zeeman broadening show little change at the radiative core
   boundary \citep[see discussion in][]{2008MNRAS.390..545D}, implying
   only a structural change in the magnetic field, possibly due to the
   theorized change in dynamo mechanism.  Additionally, for fully
   convective stars these large scale structures were observed to be
   stable over $\approx 1$yr with stability over only a few months for
   the magnetic structures of stars with radiative cores.

   Further work using Stokes I observations, which is sensitive to the
   total magnetic field, in \cite{2009arXiv0901.1659R} has similarly
   observed a sharp change in apparent field structure over the
   transition from fully convective stars to those with a radiative
   core.  \cite{2009arXiv0901.1659R} find, by comparing Stokes V and I
   components, that across this boundary (i.e. the RC gap) the
   percentage of total flux stored in large-scale structures drops
   from 15 to 6\%, with the presence of a radiative core. The mean
   total flux does not show a comparable jump. Interestingly,
   \cite{2009arXiv0901.1659R} show that 85 percent of the flux is not
   detected in Stokes V and is therefore is present in small scale
   strucures. Additionally, they note that the weak-field assumption
   and least-squares deconvolution method applied in
   \cite{2008MNRAS.390..545D} and \cite{2008MNRAS.390..567M} could
   lead to large scale fields of several kG being missed. This may
   mean that the change in magnetic field structure with the
   transition from full convective to radiative core could simply be
   due to an increase in field strength of the large-scale structures
   over the detectable threshold. This scenario is currently
   untestable, but is unlikely given our observations. Our data
   indicate a change in photometric variability and therefore magnetic
   field strucure with the presence of a radiative core. If this is
   the same mechanism as is postulated for the older field stars, and
   is indeed the result of a transition in stellar structure, then the
   change in variability observed across the RC gap arises from a
   fundamental change in magnetic field topology, itself a consequence
   of a change in the dynamo mechanism.
   
   \subsection{Implications for future variability
     surveys} \label{subsection:surveys}
 
   Variability will be an important technique for identifying young
   stellar associations in large area variability surveys such as the
   Large Synoptic Survey Telescope
   \citep[e.g.][]{2007prpl.conf..345B}.  Fig.\,\ref{fig:10arc_cmds}
   shows what such a survey should discover. Only the sequence below
   the RC gap is visible in the variability data, which reduces the
   chances of finding the cluster. However the variability data
   clearly show where the RC gap begins. Since the position of the
   RC gap fades with age, measuring this position provides a new
   technique for measuring the age of a group, which will be useful
   well beyond 20 Myr. This contrasts sharply with the results of
   \citet{mayne07_empirical_isochrones}, who concluded that the
   distance across the RC gap would become unusable as an age
   indicator for groups older than h Per, since the gap closes to be
   extremely small. Thus it appears that although the RC gap in CMDs
   alone may no longer provide ages for these older systems, a CMD of
   the variable stars could do so.

\section{Conclusion} \label{Section:Conclusion}

In conclusion we have measured, and proved the statistical significance of, a
reduction in the number of variable stars, compared to non-variables, as one
moves from below to above the RC gap in h Per. We have discussed the connection
between this reduction in variability acoss the gap and underlying changes in
the stellar magnetic fields. We suggest that this distinct reduction in
variability could be due to a change in the magnetic field topology caused by a
change in stellar structure, from fully convective stars to those with radiative
cores. This idea implies that the fully convective stars retain large-scale
stable surface magnetic fields, which lead to long-lived cool spots on the
stellar surface, and therefore long-term variability. In contrast, for stars
with a radiative core the magentic field becomes dominated by smaller scale
structures producing more transient and less significant variability. It is
clear that variability surveys will therefore be biased towards detection of the
fainter, fully convective, stars within clusters. However, the bright edge of
the RC gap, where the variability dies out, could be used as an age indicator.

\section{Acknowledgements}
   
ESS and NJM were funded through PPARC studentships. SPL is supported
by an RCUK fellowship and PPARC grant PPA/G/S/2003/00058. The INT is
operated on the island of La Palma by the Isaac Newton Group in the
Spanish Observatorio del Roque de los Muchachos of the Institute de
Astrofisica de Canarias. We would also like to thank the referee for
an in depth review and helpful comments.

\bibliographystyle{stylesheets/mn2e}
\bibliography{ref/astronomyReferences}

\end{document}